\documentclass[a4paper,11pt]{article}
\pdfoutput=1

\usepackage{jinstpub_modified}

\title{Algebraic methods for reconstruction of coordinates in strip detectors}

\author{Igor B. Smirnov}
\affiliation{Petersburg Nuclear Physics Institute, 
National Research Center ``Kurchatov Institute'', \\
Orlova roshcha 1, Gatchina, 188300, Russia}

\emailAdd{Igor.Smirnov@cern.ch}

\abstract{Many types of detectors, 
such as cathode strip chambers and some MPGDs, 
allow us to reconstruct track positions by induced strip charges.
There are two main types of algebraic methods for this reconstruction:
center of gravity methods and little-known differential methods.
General formulas for differential methods are derived.
Their performance is compared with the performance of 
center of gravity methods.
It is shown that the differential methods are better than
the previously known center of gravity methods,
but the improved center of gravity methods provide results similar to
the results of differential methods.
}


\proceeding{7th international conference on Micro Pattern Gaseous Detectors 2022, MPGD2022\\
  December 11--16, 2022,\\
  Weizmann Institute of Science, Rehovot, Israel}

\notoc

\begin{document}

\maketitle
\flushbottom

\section{Introduction}
\label{sec:intro}

Many\let\thefootnote\relax\footnotetext{This is the Accepted Manuscript version of an article accepted for publication in Journal of Instrumentation.
  Neither SISSA Medialab Srl nor IOP Publishing Ltd is responsible for any errors or omissions in this version of the manuscript or any version derived from it.
  The Version of Record is available online at
  \texttt{https://doi.org/10.1088/1748-0221/18/06/C06001}.}
types of detectors, 
such as cathode strip chambers and some MPGDs, 
allow us to reconstruct track positions by induced strip charges.
There are two main types of algebraic methods for this reconstruction:
center of gravity methods and little known differential methods.
In the differential methods the track coordinate is a ratio of linear combinations of
strip charges with parameters constrained 
by considerations of symmetry and continuity.
The resulting formulas are elegant and effective.
They do not depend on
the common pedestal.
Only special cases of these formulas can be found in the 
literature, usually under different names.
In this work general differential formulas
are derived and tested.
In order to compare them with alternative approaches,
center of gravity methods are also considered
and improved.
Only continuous methods, that is the methods that do not produce
gaps in the coordinate distributions, are considered here.
Algebraic methods may always be useful, and they are the only choice for
very high rate experiments, for which the Maximum Likelihood Estimate
(MLE) of coordinates with the strip response function takes too much computer
time.

\section{Center of gravity methods}

The Center Of Gravity method for $n$ strip charges $q_i$ with the first strip number $j$
with subtraction of the Bias level (COGB)
was introdiced in \cite{Charpak_1979}.
In our notation it is
\begin{align}
  x_{\mathrm{er}} = \sum\limits_{i=j}^{j+n-1}
    (x_{\mathrm{s},i} - x_{\mathrm{gc}}) 
\max\{0, q_i - B\} \Biggl/
\sum\limits_{i=j}^{j+n-1}\max\{0, q_i  - B\} ,
\nonumber
\end{align}
where 
$x_{\mathrm{s},i}$ is the position of the center of the strip number $i$ (``s'' means strip),
$x_{\mathrm{gc}}$ is the coordinate of the center of this strip group (the Group Center),
 $x_{\mathrm{er}}$ is the Estimate of the Relative coordinate with respect to this group center.
The coordinates are measured here in the units of strip width.
The estimate of the absolute coordinate is $x_{\mathrm{e}} = x_{\mathrm{er}} + x_{\mathrm{gc}}$.
The bias level is $B = \alpha \sum_{i=j, j+n-1} q_i$, where $\alpha$ 
is a small non-negative constant.
The resulting $x_{\mathrm{er}}$ has systematic shift and depends on  the common pedestal.
The accuracy of this method will be better,
if one rises $\max\{0, q_i  - B\}$ to a power $p$.
This gives a second adjustable parameter.
Let us call this method the Center Of Gravity with Bias level and with Power, COGBP.
Simulations show that COGBP with smoothing of the occupancy distribution (see below)
provides very good accuracy.

In a new method, called Center Of Gravity with Weight function with Splines or COGWS,
the charge is multiplied by a modulating or
weight function $w(x)$, such that $w(x) \geqslant 0$, 
$w(x)=w(-x)$, $w(x)= 0$ for $x > t$, where $t$ halfwidth.
The generalized center of gravity is expressed by the ratio: 
\begin{eqnarray}
  R(x) = 
\sum\limits_{i=j}^{j+n-1}q_i \int\limits_{i}^{i+1} w(z - x)z \mathrm{d}z  \Biggl/
  \sum\limits_{i=j}^{j+n-1} q_i \int\limits_{i}^{i+1} w(z - x) \mathrm{d}z 
\nonumber
\end{eqnarray}
The estimate of the coordinate $x_e$ is given by the equation
$x_{\mathrm{e}} = R(x_{\mathrm{e}})$. 
$w$ is a second order spline with the continuous first derivative.
Fitted parameters are coordinates of breaking points (``knots'').
Systematic shifts are negligible and the statistical resolution is close to the best possible one.
Solutions do not depend on the common pedestal.
The method is very complex. 

\section{Differential methods}

Let us suppose here for brevity that $j=1$.
Let $x_{\mathrm{er}} = \sum\limits_{i=1}^{n}a_i q_i \biggl/ \sum\limits_{i=1}^{n}b_i q_i$ with any
parameters $a_i$ and $b_i$. Here we consider the case of $n=6$.
Assume that the maximal charge is $q_4$ and $q_4 \geqslant q_3 \geqslant q_5$
(the track is most likely between 
$(x_{\mathrm{s}, 3}+x_{\mathrm{s}, 4})/2$ and $x_{\mathrm{s}, 4}$).
Then, $x_{\mathrm{er}}$ should be zero, if $q_1=q_6$, $q_2=q_5$, $q_3=q_4$ (condition of symmetry),
and $0.5$, if $q_2=q_6$, $q_3=q_5$ (condition of continuity).
Then, $x_{\mathrm{er}}$ can be expressed by
\begin{eqnarray}
x_{\mathrm{er}} = 
\frac{ a_1 d_{1,6} + a_2 d_{2,5} + a_3 d_{3,4} }
{ 2 a_1 d_{1,2}  + 2 a_2 d_{2,3} + 2 a_3 d_{3,4} + 
  b_5 d_{5,3} + b_6 d_{6,2}  },
\ \ \ \ d_{i,j} = q_i - q_j,
\nonumber
\\
a_1 \leqslant 0, \ a_2 \leqslant 0,
a_3 < 0, 
b_5 < -2 a_2, b_6 \leqslant -2 a_1 \ .
\nonumber
\end{eqnarray}
Constraints are sufficient to have a non-zero denominator for 
$q_1 \leqslant q_6 \leqslant q_2 \leqslant q_5 \leqslant q_3 \leqslant q_4  \land
( q_3 \neq q_4 \lor q_3 \neq q_5 )$.
If the maximal charge is $q_3$ and $q_3 \geqslant q_4 \geqslant q_2$
(the track is most likely between 
$x_{\mathrm{s}, 3}$ and $(x_{\mathrm{s}, 3}+x_{\mathrm{s}, 4})/2$),
it needs to swap $q_1 \leftrightarrow q_6$, $q_2 \leftrightarrow q_5$, $q_3 \leftrightarrow q_4$,
to calculate $x_{\mathrm{er}}$ and to change its sign. Let us call this formula
Asymmetric Differential Formula, ADF.
Symmetric Differential Formula, SDF, does not require permutations:
\begin{eqnarray}
x_{\mathrm{er}} = \frac{1}{2}\ \frac{ a_1 d_{1,6} + a_2 d_{2,5} + 
  (a_2 - a_1) d_{3,4}}
{ a_1 s_{1,6} + (a_2 - 2 a_1) s_{2,5} + (a_1 - a_2) s_{3,4}},
\ \ \ \ s_{i,j} = q_i + q_j,
\ \ \ a_1 \leqslant 0,\  a_2 < a_1
\nonumber
\end{eqnarray}
Both ADF and SDF can be used with the bias and power (notation with suffix BP): 
$q_i\rightarrow (\max\{0, q_i - B\})^p$.
At $a_1 = 0$ and $b_6 = 0$ these formulas are converted in
4-strip formulas. 
The symmetric 4-strip formula
$x_{\mathrm{er}} = 0.5(-q_1-q_2+q_3+q_4)/(-q_1+q_2+q_3-q_4)$
is proposed in Ref. \cite{Spiridenkov_1994}.

Similar asymmetric and symmetric formulas can be derived for 7 strips.
If $a_1=a_2=b_6=b_7=0$, then 3-strip formulas are obtained from them.
If $b_3 = 0$, the asymmetric 3-strip formula is ``the ratio method''
from
Ref. \cite{Khovansky_1994}.
The symmetric 3-strip formula 
$x_{\mathrm{er}} =  0.5( -q_1 + q_3)/( -q_1 + 2 q_2 - q_3 )$
is also an algebraic fit of the parabolic
strip response function
\cite{Endo_1981}. 

It can be proved that 
similar asymmetric and symmetric formulas exist for any greater number of strips.

\section{Correction of systematic shifts}

Distortions of occupancy distributions can be completely corrected 
by ``smoothing''  this distribution using the cumulative distribution function
of $x_{\mathrm{er}}$ \cite{Chiba_1983,Belau_1983}.
This procedure approximately corrects the systematic shifts of $x_{\mathrm{er}}$ as well.
The COGBP and ADF methods with corrections by this procedure
allow one to provide almost the best possible precision.

\section{Conclusion}

A range of new differential methods
and center of gravity methods is developed. 
Several methods of both types applied with smoothing 
and
the COGWS method (the center of gravity method 
for which the smoothing is unnecessary)
provide almost zero systematic shifts and 
the resolution which is very close to the best resolution
attainable by MLE.
Several methods of both types
(SDF, ADF, COGWS) ensure the independence of the results
on the common pedestal.

\end{document}